\documentclass[trackchanges]{aastex701}

\begin{document}

\title{Interstellar Comet 3I/ATLAS: Evidence for Galactic Cosmic Ray Processing}

\author[orcid=0000-0002-5658-1313]{Romain Maggiolo}
\affiliation{Royal Belgian Institute for Space Aeronomy (BIRA-IASB) Brussels, Belgium}
\email[show]{romain.maggiolo@aeronomie.be}  

\author[orcid=0000-0001-7334-833X,gname=Bosque, sname='Sur America']{Frederik Dhooghe} 
\affiliation{Royal Belgian Institute for Space Aeronomy (BIRA-IASB) Brussels, Belgium}
\email{frederik.dhooghe@aeronomie.be}

\author[orcid=0000-0002-0331-7076]{Guillaume P. Gronoff}
\affiliation{Science Systems and Application Inc. Hampton, Va, USA}
\affiliation{Nasa Langley Research Center, Hampton, Va, USA}
\email{guillaume.p.gronoff@nasa.gov}

\author[orcid=0000-0003-4805-5695]{Johan de Keyser}
\affiliation{Royal Belgian Institute for Space Aeronomy (BIRA-IASB) Brussels, Belgium}
\affiliation{Centre for mathematical Plasma Astrophysics, Heverlee, Belgium}
\email{johan.dekeyser@aeronomie.be}

\author[orcid=0000-0002-4316-7252]{Gaël Cessateur}
\affiliation{Royal Belgian Institute for Space Aeronomy (BIRA-IASB) Brussels, Belgium}
\email{gael.cessateur@aeronomie.be}



\begin{abstract}

Spectral observations of 3I/ATLAS (C/2025~N1) with \textit{JWST}/NIRSpec and \textit{SPHEREx} reveal an extreme CO$_2$ enrichment 
(\mbox{CO$_2$/H$_2$O = 7.6$\pm$0.3}) that is 4.5~$\sigma$ above solar system comet trends and among the highest ever recorded. 
This unprecedented composition, combined with substantial absolute CO levels 
(\mbox{CO/H$_2$O = 1.65$\pm$0.09}) and red spectral slopes, provides direct evidence for 
galactic cosmic ray (GCR) processing of the outer layers of the interstellar comet nucleus. 

Laboratory experiments demonstrate that GCR irradiation efficiently converts CO to CO$_2$ 
while synthesizing organic-rich crusts, suggesting  that the outer layers of 3I/ATLAS consist of irradiated material which properties are consistent with the observed composition of 3I/ATLAS coma and with its observed spectral reddening. 
Estimates of the erosion rate of 3I/ATLAS indicate that current outgassing samples the 
GCR-processed zone only (depth $\sim$15--20~m), never reaching pristine interior material. 
Outgassing of pristine material after perihelion remains possible, though it is considered unlikely. 

This represents a paradigm shift: long-residence interstellar objects primarily reveal 
GCR-processed material rather than pristine material representative of their primordial 
formation environments. With 3I/ATLAS approaching perihelion in October~2025, immediate 
follow-up observations are critical to confirm this interpretation and establish GCR 
processing as a fundamental evolutionary pathway for interstellar objects.

\end{abstract}



\section{Introduction} 

The discovery of interstellar objects 1I/`Oumuamua and 2I/Borisov has opened a new window 
into the composition and evolution of small bodies from other stellar systems. 
1I/`Oumuamua exhibited puzzling non-gravitational acceleration despite showing no detectable outgassing, 
suggesting either exotic composition or unusual surface properties 
\citep{Micheli2018,Seligman2020}. 
2I/Borisov displayed more conventional cometary activity but with unusually high CO enrichment 
relative to solar system comets, indicating potential differences in formation environment or 
evolutionary history \citep{Bodewits2020,Cordiner2020}.

A fundamental question for understanding interstellar objects concerns how extended residence 
in the galactic environment affects their composition and surface properties. 
Dose deposition models for solar system comets over Gyr timescales \citep{Gronoff2020}, 
combined with laboratory experiments on cometary ice analogs 
\citep{Johnson1997,Hudson1999,Dartois2015}, 
show that galactic cosmic ray (GCR) irradiation can substantially alter nucleus material over 
Gyr time scales, modifying its original chemical composition and ice structure 
\citep{Maggiolo2020}. 
However, no direct observational evidence for such processing has been identified in confirmed 
interstellar visitors.

The interstellar comet 3I/ATLAS (C/2025~N1) was discovered on 2025~July~1 by the 
Asteroid Terrestrial-impact Last Alert System (ATLAS; \citealt{Tonry2018}) from Rio Hurtado, Chile. 
Its interstellar nature was confirmed the following day \citep{Denneau2025}. 
This third interstellar object provides a new opportunity to investigate these questions. 
The object follows a strongly hyperbolic trajectory 
($e \simeq 6.1$, $q \simeq 1.36$~AU, $i \simeq 175^{\circ}$, $V_{\infty} \simeq 58$~km~s$^{-1}$), 
confirming its extrasolar origin \citep{Seligman2025,Bolin2025}. 
High-resolution imaging constrains the effective nucleus radius to 
$r \lesssim 2.8$~km, with volatile-dependent lower bounds of 
$r \gtrsim 0.22$--1.1~km derived from coma properties \citep{Jewitt2025}.

Dynamical analyses of 3I/ATLAS suggest it is the oldest known interstellar object, 
with kinematic models estimating its age between 3 and 11~Gyr \citep{Taylor2025}. 
galactic orbit comparisons indicate it likely originated from an ancient, 
metal-poor stellar population older than 7.6~Gyr \citep{Hopkins2025}. 
Early spectroscopic observations reveal active outgassing with compositional signatures 
that differ markedly from both previous interstellar objects and solar system comets.

In this communication, we discuss spectroscopic observations of 3I/ATLAS within the 
framework of cosmic ray processing mechanisms. 
We demonstrate that the observed signatures provide compelling evidence for 
Gyr-scale galactic cosmic ray alteration of the outer layers of 3I/ATLAS's nucleus, 
with implications for interpreting the composition of ancient visitors from other stellar systems.

\section{Observations} \label{sec:style}

\subsection{Spectroscopic Measurements} \label{sec:spectroscopy}

Near-nucleus \textit{JWST}/NIRSpec Integral Field Unit (IFU) spectra obtained on 
UT~2025~August~6 at a heliocentric distance of $r_{\mathrm{h}} \simeq 3.32$~au 
reveal a coma dominated by CO$_2$ outgassing with detectable CO and H$_2$O production 
close to the nucleus. 
From the nucleus-centered extract, \citet{Cordiner2025} derive production rates of 
$Q(\mathrm{CO_2}) = (9.50\pm0.08)\times10^{26}$~s$^{-1}$, 
$Q(\mathrm{CO}) = (1.70\pm0.04)\times10^{26}$~s$^{-1}$, and 
$Q(\mathrm{H_2O}) = (1.07\pm0.07)\times10^{26}$~s$^{-1}$. 
Based on their $Q$-curve analysis, \citet{Cordiner2025} estimate the total 
coma-equivalent gas production rates for 3I/ATLAS. 
The $Q$-curves show CO$_2$ and CO flatten by $\sim$3000~km, indicating that gas production 
for these species is confined within $\sim$3000~km of the nucleus. 
The H$_2$O $Q$-curve shows no clear asymptote, providing evidence for an extended water source. 

The resulting terminal gas production rates are 
$Q(\mathrm{CO_2}) = (1.70\pm0.01)\times10^{27}$~s$^{-1}$, 
$Q(\mathrm{CO}) = (3.7\pm0.2)\times10^{26}$~s$^{-1}$, 
$Q(\mathrm{H_2O}) = (2.23\pm0.08)\times10^{26}$~s$^{-1}$, and 
$Q(\mathrm{OCS}) = (1.7\pm0.9)\times10^{24}$~s$^{-1}$. 
Based on the $Q$-curve terminal values, the coma-averaged mixing ratios are 
CO$_2$/H$_2$O~=~$7.6\pm0.3$ and CO/H$_2$O~=~$1.65\pm0.09$.

Complementary \textit{SPHEREx} measurements (2025~August~12, 
$r_{\mathrm{h}} \simeq 3.1$--3.3~au) find 
$Q(\mathrm{CO_2}) \approx 9.4\times10^{26}$~s$^{-1}$, and 3$\sigma$ upper limits of 
$Q(\mathrm{H_2O}) \le 1.5\times10^{26}$~s$^{-1}$ and 
$Q(\mathrm{CO}) \le 2.8\times10^{26}$~s$^{-1}$. 
These are qualitatively consistent with JWST’s CO$_2$ dominance and weaker H$_2$O/CO near the 
nucleus. Quantitatively, JWST’s terminal H$_2$O production rate 
($2.19\times10^{26}$~s$^{-1}$) is modestly above the \textit{SPHEREx} 3$\sigma$ limit, 
which is reasonable given the different apertures and sensitivities 
\citep{Lisse2025}. 

\textit{Swift}/UVOT OH imaging yields 
$Q(\mathrm{H_2O}) = (0.74\pm0.50)\times10^{27}$~s$^{-1}$ at 
$r_{\mathrm{h}}=3.51$~au (2025~July~31) and 
$(1.36\pm0.35)\times10^{27}$~s$^{-1}$ at $r_{\mathrm{h}}=2.90$~au (2025~August~19), 
measured in a $10''$-radius ($\simeq 2.0\times10^{4}$~km) aperture 
\citep{Xing2025}. 
The larger-aperture \textit{Swift} rates, together with JWST’s outward-increasing 
H$_2$O $Q$-curve, support an extended water source.

The color and spectral properties of 3I/ATLAS have been characterized through photometric 
and spectroscopic observations spanning the visible to near-infrared. 
3I/ATLAS consistently exhibits a steep, red spectral slope and color indices indicative 
of a red continuum. 
Spectrophotometry yields color indices of $g-r \simeq 0.60$ and $r-i \simeq 0.21$, 
consistent with its red reflectance spectrum \citep{Beniyama2025}. 
Photometry yields color indices such as $B-V = 0.98\pm0.23$, 
$V-R = 0.71\pm0.09$, and $g-r = 0.84\pm0.05$, corresponding to a reflectance slope of 
$(16.0\pm1.9)\%/100$~nm \citep{Bolin2025}. 
Spectrophotometric measurements corroborate these findings, reporting slopes of 
$\sim$19\%/100~nm over 420--700~nm \citep{Belyakov2025}, 
$18.3\pm0.9\%/100$~nm across 400--900~nm \citep{MarcosDeLaFuente2025}, and 
$17.1\pm0.2\%/100$~nm across 420--700~nm \citep{Seligman2025}. 
In the near-infrared, the slope flattens, trending toward neutral or slightly blue 
reflectance beyond $\sim$1.1~$\mu$m \citep{Belyakov2025,Kareta2025}. 
Despite broad agreement on its red spectral character, precise interpretation is hindered 
by coma dominance, instrumental systematics, and poorly constrained dust properties 
\citep[e.g.,][]{Seligman2025,Kareta2025}.

\subsection{Comparison with Known Populations} \label{sec:comparison}

The CO$_2$/H$_2$O ratio of $7.6\pm0.3$ places 3I/ATLAS among the most CO$_2$-enriched 
objects observed. Compared to solar system comets, the comprehensive survey of 
\citet{HarringtonPinto2022} reported a median CO$_2$/H$_2$O ratio of 
$0.12\pm0.02$ across 25~comets, with a few outliers reaching up to, but not exceeding, 
$\sim$0.3 in individual cases \citep{HarringtonPinto2022}, consistent with the upper 
values reported in earlier surveys \citep{Ootsubo2012,McKay2015}. 
The statistical significance places the CO$_2$/H$_2$O ratio 4.5$\sigma$ above solar system 
trends \citep{Cordiner2025}, although this enrichment must be interpreted cautiously 
given observational selection effects and the limited sample of comets observed at 
comparable heliocentric distances where CO$_2$ can be measured. This ratio also significantly exceeds interstellar medium CO$_2$/H$_2$O ice abundances, 
typically $\sim$10--30\% with environment-dependent maxima up to $\sim$40--50\%, as summarized 
by \citet{Boogert2015} and confirmed in JWST background-star sightlines 
($\sim$10--20\%) by \citet{McClure2023}. 

The high CO/H$_2$O ratio ($1.65\pm0.09$) also greatly exceeds the solar system median of 
$0.03\pm0.01$ \citep{HarringtonPinto2022}. 
Most comets inside $\sim$3~au span 0.002--0.23 \citep{BockeleeMorvan2017}. 
This enrichment is similar to 2I/Borisov, which showed 
CO/H$_2$O~$\approx$~1.3--1.55 near perihelion \citep{Bodewits2020}, suggesting possible 
commonalities among interstellar objects that may reflect shared formation or evolutionary 
processes.

The spectral slope of 3I/ATLAS places it among the reddest small bodies known in the 
solar system. Its visible continuum is considerably steeper than that observed in typical 
cometary comae \citep{Opitom2025}, yet is comparable to the slopes measured for interstellar 
objects such as 1I/`Oumuamua \citep{Fitzsimmons2018} and 2I/Borisov \citep{deLeon2020}. 
Within solar system reservoirs, 3I/ATLAS overlaps with the redder members of the Centaur 
population \citep{Fornasier2009}, including very red objects such as Pholus, and is consistent 
with the spectrophotometry of trans-Neptunian objects (TNOs) such as 2012~DR30, whose 
reflectance spectrum exhibits a pronounced reddening in the near-UV \citep{Seccull2021}. 
3I/ATLAS is less red than the ``ultrared'' cold classical TNOs, which display the steepest 
optical slopes among trans-Neptunian populations, as shown by the \textit{HST}/WFC3 survey \citep{Fraser2012} and more recent JWST observations \citep{SouzaFeliciano2024}. 
Many of these objects also exhibit deviations from a strictly linear reddening trend in the 
near-infrared \citep{Seccull2021}. 
Its slope also falls within the range of D-type asteroids, known for their steep, red, 
featureless spectra \citep{Fitzsimmons1994}.

\section{Evidence for cosmic ray processing if 3I/Atlas} \subsection{Primordial and Post-Formation Pathways for 3I/ATLAS's Observed Properties} \label{sec:pathways}

\textit{JWST}/NIRSpec measurements show CO$_2$/H$_2$O $= 7.6 \pm 0.3$, placing 3I/ATLAS
$4.5\sigma$ above solar system trends at the observed heliocentric distance. Any viable
explanation must therefore elevate the near-surface CO$_2$/H$_2$O ratio substantially.

Multiple processes could contribute to the unusual volatile inventory of comets. Physical
conditions during the formation of protoplanetary discs \citep[e.g.,][]{Drozdovskaya2016}
and chemical evolution within protoplanetary disks can alter CO and CO$_2$ abundances
depending on disk conditions \citep[e.g.,][]{Eistrup2018}. In addition, non-thermal
processing of ices by GCR and UV photons in dense molecular clouds
and protoplanetary disks can modify ice composition before incorporation into planetesimals,
thereby influencing the volatile make-up inherited by comets \citep{Semenov2004}. Pebble
drift mechanisms can yield CO/H$_2$O $\gtrsim 10$ by creating CO-rich rings
\citep{Price2021}, but because CO$_2$ condenses closer to the star than CO, their
snowlines are spatially separated, preventing simultaneous enhancement of both species in
the same reservoir. Additional disk pathways, such as midplane CO destruction that converts
CO into CO$_2$ and other species \citep{Bosman2018}, or clathrate trapping and
amorphous-ice inheritance affecting volatile budgets \citep{Mousis2010}, can alter
carbon–oxygen partitioning but likewise struggle to reproduce \emph{simultaneously}
CO$_2$- and CO-rich ices at the observed level. Statistical comparisons of disk chemical
evolution models with cometary observations confirm partial agreement but reveal persistent
mismatches in CO/CO$_2$ ratios, even when formation is assumed near the CO snowline
\citep{Eistrup2019}. Taken together, these primordial scenarios each reproduce certain
aspects of cometary volatile inventories but struggle to account for the simultaneous
presence of both CO- and CO$_2$-rich ices. This motivates consideration of post-formation
processes that could further modify initial compositions. GCR processing merits particular
attention given indications that 3I/ATLAS may be an ancient interstellar object and the
well-documented effects of energetic particle irradiation on ice composition.

Experimental results consistently demonstrate that irradiation of H$_2$O--CO and
H$_2$O--CO$_2$ ices by energetic particles leads to CO$_2$ formation. This process was
identified in ion-irradiation studies \citep{Palumbo1997} and was subsequently confirmed
through proton bombardment of mixed ices \citep{Hudson1999}. These findings have been
extended across a broad range of ice compositions, temperatures, and radiation sources,
including both ions and ultraviolet photons \citep{Gerakines2001}. Cold H$_2$O--CO ices,
analogous to those found in dense molecular clouds and cometary environments, reliably show
efficient CO$_2$ synthesis, with yields modulated by ice thickness, mixing ratio,
projectile energy, and temperature \citep{Gerakines2001,Mennella2004}. Furthermore, CO$_2$
formation occurs readily from various C--O-bearing ice mixtures
\citep[e.g.,][]{Palumbo1997,Baratta2002}. Despite the consistent production of CO$_2$, CO
is not entirely consumed; multiple studies report that CO persists in the irradiated ice
\citep[e.g.,][]{Palumbo1997,Satorre2000,Mennella2004}. Notably, ion-irradiation experiments
involving CO$_2$, either as pure ice or in mixtures, show that the CO$_2$/CO ratio
stabilizes at approximately 0.1 after a fluence of $\sim 2$--$3 \times 10^{12}$~ions
cm$^{-2}$, independent of the initial CO$_2$/H$_2$O ratio \citep{Pilling2010}. This
persistence is attributed to the degradation of complex organic molecules, which are
synthesized from simple species during irradiation and subsequently broken down into smaller
fragments, including CO and CO$_2$ \citep{Gerakines2001}. This feedback mechanism creates a
dynamic CO--CO$_2$ equilibrium environment in which CO$_2$ dominates the volatile
inventory, while CO is continuously regenerated from organic decomposition. While laboratory
experiments are typically conducted over short timescales and at high irradiation fluxes
with relatively simple ice mixtures, making direct extrapolation to gigayear-scale
astrophysical environments uncertain, the fundamental radiolysis pathways leading to CO$_2$
enrichment and CO persistence are robust and well characterized.

Monte Carlo radiation transport simulations with the \textsc{CometCosmic} code, built on
\textsc{Geant4}, show that GCR are the dominant long-term radiation source in cometary
nuclei \citep{Gronoff2020}. Energy deposition is controlled mainly by primary protons and
their secondary electromagnetic cascades. The deposited dose decreases with depth but remains
significant to 15--20~m, with detectable deposition extending to $\sim$100~m. Short-lived
supernova cosmic-ray events can increase the dose by orders of magnitude, but only for
timescales of $10^{4}$--$10^{5}$~yr and thus contribute negligibly to the integrated
Gyr-scale budget compared to steady GCR irradiation.

Building on the dose deposition results, \citet{Maggiolo2020} show that over 4.5~Gyr of GCR
irradiation, the outer 10--20~m of a comet nucleus similar to 67P would undergo chemical
and structural changes. To translate the dose deposition into compositional effects, they
applied laboratory radiolysis yields from H$_2$O--CO mixtures \citep{Hudson1999} and pure
H$_2$O ice \citep{Johnson1997}, predicting the formation of O$_2$, H$_2$O$_2$, and CO$_2$,
with CO$_2$ produced effectively in H$_2$O--CO ice mixtures. Using the \citet{Hudson1999}
yield, they find that essentially all CO initially present in H$_2$O--CO ices can be
converted into CO$_2$ within the first $\sim$10~m of the surface after 4.5~Gyr of GCR
exposure. In addition to CO$_2$, the model predicts the accumulation of O$_2$ and H$_2$O$_2$
in the irradiated layers. Because molecular diffusion is limited at cometary temperatures,
these products remain confined to a shallow processed zone. Irradiation also induces
amorphization and compaction of ice layers, modifying the physical structure of the crust.
\citet{Maggiolo2020} further evidence the production of $^{15}$N from GCR-induced spallation
of $^{16}$O, where neutrons generated by high-energy particle impacts
(1~MeV--1~TeV~nucleon$^{-1}$) are captured to form heavier nuclei, yielding $^{15}$N mainly
through the reaction n + $^{16}$O $\rightarrow$ $^{15}$N + d + e$^{-}$. The production of
$^{15}$N remains very low, with a maximum increase of the $^{15}$N/$^{14}$N ratio of
$\sim 2 \times 10^{-4}$ after 4.5~Gyr of irradiation by GCR. This is about a factor of 25
too low to explain the high $^{15}$N/$^{14}$N ratio observed in comets \citep[see][]{Gronoff2020}.
However, it could be significant for a body with a low initial $^{15}$N/$^{14}$N ratio.\\
Energetic particle irradiation of icy surfaces, such as those found on comets and trans-Neptunian objects, drives the radiolytic transformation of simple volatiles (e.g., H$_2$O, CH$_4$, NH$_3$, CO) into complex, refractory organic residues. This process begins with molecular dissociation and radical formation, followed by recombination into larger macromolecules, ultimately producing a carbon-rich organic crust \citep{Strazzulla2001}. Repeated irradiation increases the complexity of the crust, resulting in surface darkening and spectral reddening. This arises from the preferential destruction of high-albedo ices and the accumulation of organic molecules, which absorb efficiently in the blue and reflect more in the red to near-infrared wavelengths resulting in strong reddening and darkening of spectral reflectance similar to that seen on Centaurs and trans-Neptunian objects \citep{Brunetto2006}. The reddening effect depends not only on radiation dose but also on irradiation temperature, explaining some of the diversity in TNO colors \citep{Zhang2025}\\
\citet{Merkulova2025} suggest that GCR processing of cometary nuclei could trigger explosive
outbursts during modest heating events (associated, for instance, with a passing star,
galactic tide, or cosmic dust impact). They propose that irradiated ice in cometary nuclei
gradually accumulates energetic free radicals (e.g., H, OH). Heating events could trigger
rapid recombination, releasing heat, converting amorphous ice to crystalline form, building
subsurface pressure, and eventually producing outbursts. Such outbursts could explain the
non-gravitational motion of some long-period comets or interstellar objects.

As comets or interstellar objects approach a star, solar heating alters their nuclei.
Thermal-evolution models of repeated perihelion passages indicate that CO is depleted more
rapidly than CO$_2$ or H$_2$O, while gas escape leads to the accumulation of refractory
surface mantles \citep{Prialnik1988}. These mantles reduce outgassing and enhance thermal
insulation, thereby steepening temperature gradients between surface and subsurface layers
\citep{GuilbertLepoutre2015}. Mantle formation is therefore widely invoked in comet evolution
models and is consistent with the observed suppression of activity in many nuclei
\citep{BockeleeMorvan2017}. This framework provides a natural explanation for the observed
dichotomy: Jupiter-family comets, which have undergone repeated solar processing, typically
display CO/H$_2$O ratios below a few percent \citep{Ootsubo2012}, whereas dynamically new
Oort-cloud comets can retain CO at levels of several tens of percent, consistent with
long-term cryogenic storage prior to their first perihelion passage \citep{McKay2015}. A
key limitation, however, is that once mantles become sufficiently thick, they suppress the
fluxes of both CO and CO$_2$, inconsistent with ATLAS's strong CO$_2$ production
\citep{GuilbertLepoutre2015}. Thermal evolution may have contributed to the present properties
of ATLAS but cannot alone account for its observed volatile ratios.

Thermal processing prior to 3I/ATLAS entry in the solar system cannot be ruled out from
current observations. Assessing its contribution is beyond the scope of this paper, which
instead focuses on GCR processing. Indeed, unlike thermal evolution, which requires close
stellar passages, either in the system where it formed or during subsequent stellar
encounters on its way to the solar system, irradiation by GCR is unavoidable. The only
uncertainty is the exposure time, although age estimates for 3I/ATLAS point toward several
billion years of irradiation.

\subsection{GCR Processing Depth} \label{sec:gcrdepth}

To estimate the penetration depth of GCR in the nucleus of 3I/ATLAS, we consider
continuous irradiation over 1~Gyr under a flux representative of the interstellar
environment of the Oort Cloud, following the approach of \citet{Gronoff2020}. The GCR flux
is calculated using the modified Badhwar and O'Neill (1992) H-BON10 model
\citep{Mertens2013}, which computes ion fluxes across all species. We consider no modulation
of the GCR flux by the heliospheric magnetic field, consistent with the conditions expected
in the interstellar medium far from the Sun. The assumption of a constant GCR flux is
supported by \citet{Smith2019}, who analyzed cosmogenic nuclides (e.g., $^{41}$K, $^{36}$Cl,
and noble gases) in iron meteorites and demonstrated that the GCR intensity
has remained stable within $\pm$10--20\% over at least the past 1~Gyr.

We adopt a bulk density of 500~kg~m$^{-3}$ as representative of cometary nuclei, consistent
with spacecraft measurements of comet 67P/Churyumov--Gerasimenko, which yielded
$\sim$530~kg~m$^{-3}$ \citep{Paetzold2016}, and with broader estimates for the comet
population derived from remote sensing and modelling \citep{Weissman2008}. We assume a
nucleus composed of water ice (of isotopic composition $^{16}$O and $^{1}$H) and SiO$_2$
with a $\mathrm{SiO_2}/\mathrm{H_2O}$ mass ratio of 4, as in \citet{Gronoff2020}.

Figure~1 shows the corresponding energy dose deposited by GCR after 1~Gyr of irradiation.
The deposited energy peaks at the surface, reaching $\sim 5\times 10^{23}$~eV~cm$^{-3}$, and
decreases with depth into the nucleus to $\sim 5\times 10^{22}$~eV~cm$^{-3}$ at 10~m and
$\sim 3\times 10^{19}$~eV~cm$^{-3}$ at 30~m.

From the energy deposition we evaluate the cumulative effects of GCR exposure on both ice
structure and chemical composition after 1~Gyr of irradiation (Figures~2 and~3), following
the same approach as in \citet{Maggiolo2020} to simulate the effect of GCR irradiation on
the nucleus of comet 67P. While these results are based on models and laboratory data that
may not fully capture real cometary conditions, they provide a robust indicator of
processing depth. The simulations show that 1~Gyr of GCR irradiation is sufficient to
significantly alter the physical state of the ice, converting it into compact amorphous
ice, and to produce secondary chemical species (CO$_2$, CH$_4$, H$_2$CO, CH$_3$OH, HCO)
from a CO--H$_2$O ice mixture down to depths of approximately 15--20~m.

\begin{figure}[ht!]
\centering
\includegraphics[width=\textwidth]{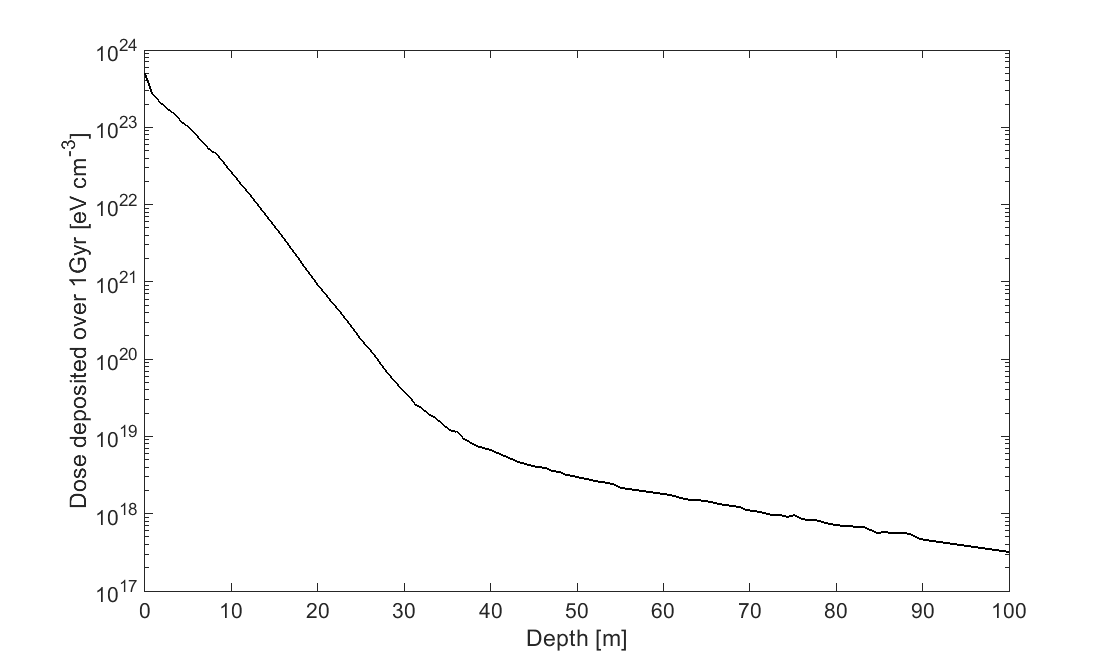}
\caption{
Energy dose deposited by GCRs in a nucleus after an exposure time of 1~Gyr in the
local interstellar medium, adapted from \citet{Gronoff2020}. The simulation assumes a bulk
density of 0.5~g\,cm$^{-3}$ and a nucleus composed of H$_2$O (of isotopic composition
$^{16}$O and $^{1}$H) and SiO$_2$ with a SiO$_2$/H$_2$O mass ratio of~4.
}
\label{fig:dose}
\end{figure}

\begin{figure}[ht!]
\centering
\includegraphics[width=\textwidth]{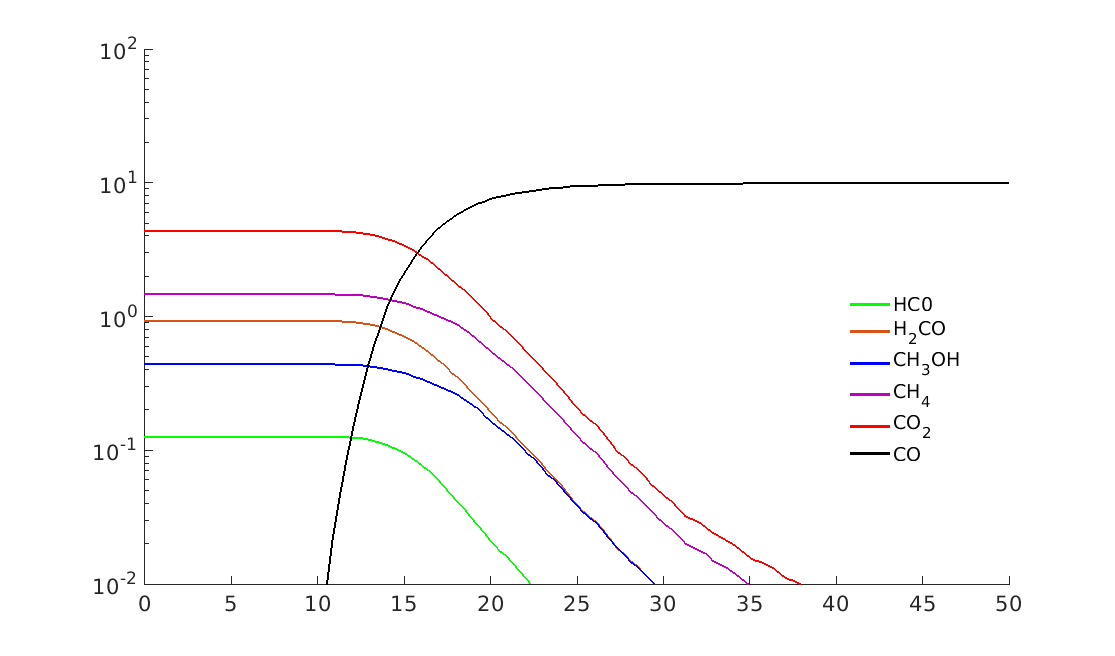}
\caption{
Depth profiles of chemical abundances relative to H$_2$O in a nucleus irradiated by
galactic cosmic rays, adapted from \citet{Maggiolo2020}. The simulation assumes a bulk
density of 0.5~g\,cm$^{-3}$, an exposure time of 1~Gyr, and mixed H$_2$O--CO ices.
Radiolysis yields are taken from laboratory experiments by \citet{Hudson1999}, while the
depth-dependent energy deposition is based on Monte Carlo radiation transport results from
\citet{Gronoff2020}. CO (black) is progressively converted into CO$_2$ (red) within the
outer $\sim$15--20~m, while other products (CH$_4$, CH$_3$OH, H$_2$CO, HCO) decrease with
depth as irradiation effects diminish.
}
\label{fig:chemical}
\end{figure}

\begin{figure}[ht!]
\centering
\includegraphics[width=\textwidth]{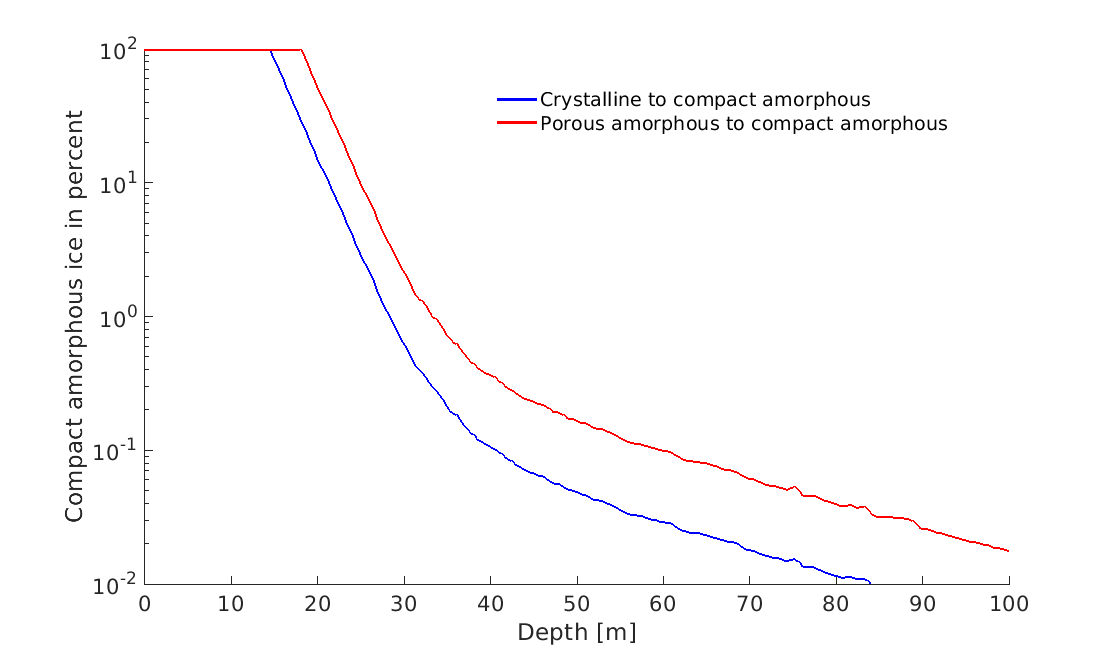}
\caption{
Depth-dependent transformation of ice structure under galactic cosmic rays irradiation,
adapted from \citet{Maggiolo2020}. The simulation assumes a bulk density of
0.5~g\,cm$^{-3}$ and an exposure time of 1~Gyr, with energy deposition rates from
\citet{Gronoff2020}. Transformation yields from \citet{Dartois2013} for porous amorphous
ice and \citet{Dartois2015} for crystalline ice are applied to compute the compact
amorphous fraction (in~\%), showing the progressive conversion of crystalline H$_2$O ice
(blue) and porous amorphous ice (red). The conversion is most efficient within the outer
tens of meters and decreases with depth.
}
\label{fig:ice}
\end{figure}

\subsection{Erosion Rate Calculation} \label{sec:erosion}

A key quantitative test of the GCR processing hypothesis is whether the depth of surface
erosion during 3I/ATLAS's Solar system passage is small compared to the expected depth of
GCR-induced alteration. To compute the erosion of 3I/ATLAS during its solar system
crossing, we model the cumulative erosion depth due to volatile outgassing and dust
ejection. The method assumes a time-dependent total mass-loss rate, $Q(t)$, derived from
available observations of gas and dust activity.

Cometary production rates are typically described by power-law dependencies on heliocentric
distance, with the slope ($\alpha$) reflecting the dominant sublimating species. For
H$_2$O-driven activity, $\alpha$ values near $-2$ are common under equilibrium conditions
\citep[e.g.,][]{Swamy1991}, with steeper slopes up to $-4$ observed near perihelion, as in
67P/Churyumov--Gerasimenko \citep{Skorov2020}. However, at heliocentric distances beyond
$\sim$3~au, where water sublimation is inefficient, the activity is more likely driven by
hypervolatiles such as CO and CO$_2$, for which survey-based fits suggest shallower
exponents: $\alpha \approx -1$ to $-2$ for CO \citep{Womack2017} and
$\alpha \approx -2$ to $-3$ for CO$_2$ \citep{Ootsubo2012}, while H$_2$O-driven activity
typically follows $\alpha \approx -3$ to $-5$ within $\sim$3~au \citep{Combi2019}. Therefore,
the commonly cited $\alpha=-2$ and $\alpha=-4$ slopes span the expected range for different
volatiles, though care must be taken when interpreting them in the context of species-specific
drivers. For 3I/ATLAS, we adopt whole-coma production rates from \citet{Cordiner2025},
$Q(\mathrm{CO_2}) = (1.76 \pm 0.02)\times 10^{27}$~s$^{-1}$,
$Q(\mathrm{CO}) = (3.0 \pm 0.2)\times 10^{26}$~s$^{-1}$,
$Q(\mathrm{H_2O}) = (2.19 \pm 0.08)\times 10^{26}$~s$^{-1}$, and
$Q(\mathrm{OCS}) = (4.3 \pm 0.9)\times 10^{24}$~s$^{-1}$. These volatiles collectively
correspond to a total gas mass-loss rate of $\sim$150~kg~s$^{-1}$ at 3.32~au. Dust
mass-loss estimates are taken from \citet{Jewitt2025}, who reported 12--120~kg~s$^{-1}$ at
3.83~au. We therefore consider two bounding cases: a ``minimum'' total mass-loss scenario
including 12~kg~s$^{-1}$ of dust, and a ``maximum'' scenario including 120~kg~s$^{-1}$ of
dust. To capture the uncertainty in heliocentric scaling, we evaluate both
$Q \propto r_{\mathrm{h}}^{-2}$ and $Q \propto r_{\mathrm{h}}^{-4}$ dependencies, representing
relatively shallow and steep activity laws, respectively.

To translate the mass-loss rate into physical erosion of the nucleus, we consider two
possible nucleus radii for 3I/ATLAS, following the size estimates reported by
\citet{Jewitt2025}: 500~m and 2200~m. These values bracket the plausible size range and
allow us to assess how erosion depth scales with nucleus size. We use a bulk density of
500~kg~m$^{-3}$, consistent with the value adopted in Section~\ref{sec:gcrdepth}.

The erosion depth as a function of time is shown in Figure~4. The heliocentric distance of
3I/ATLAS is derived from its current orbital parameters. The results indicate that material
loss due to sublimation and dust ejection remains limited before perihelion
($q = 1.356$~au on 2025-10-29 11{:}44~UT), with the bulk of the erosion occurring in the
vicinity of perihelion, where solar insolation, and hence mass loss, is the highest.
Pre-perihelion erosion remains shallow, below $\sim$1~m. Near perihelion, the erosion rate
rises sharply, leading to total erosion depths of several tens of meters in strong
outgassing scenarios. In contrast, for low and moderate outgassing scenarios, post-perihelion
erosion remains well below the thickness of the GCR-processed crust (15--20~m).

\begin{figure}[ht!]
\centering
\includegraphics[width=\textwidth]{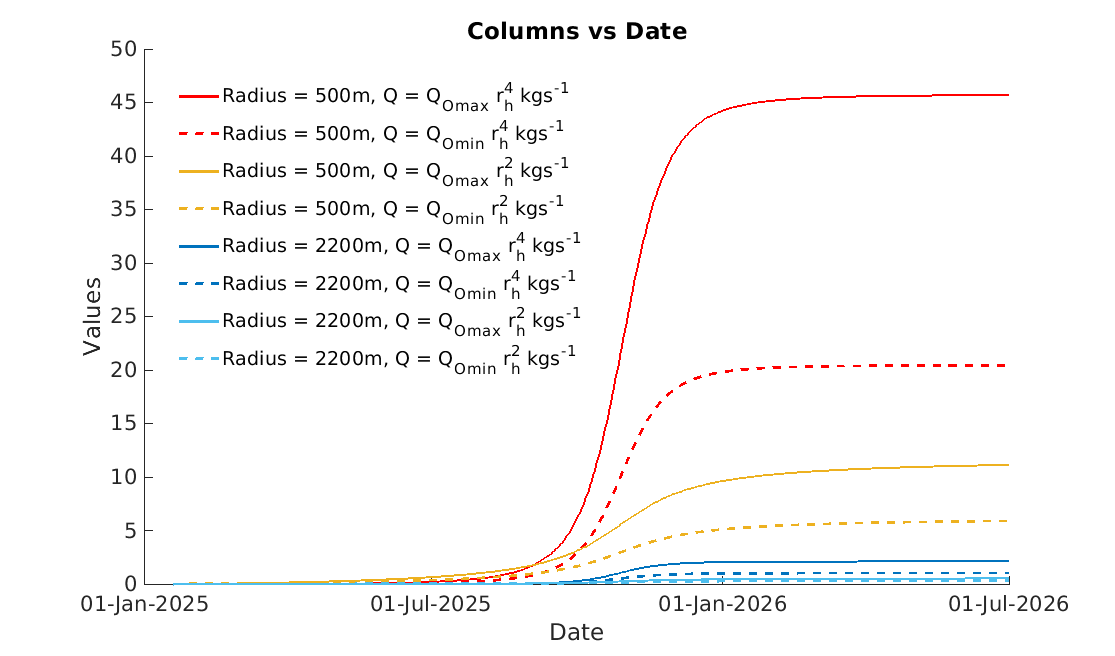}
\caption{
Cumulative erosion depth of 3I/ATLAS as a function of time during its solar system
passage, scaled with heliocentric distance from the object’s orbital parameters.
Mass-loss rates are modeled as power laws in heliocentric distance ($Q \propto r_{\mathrm{h}}^{-2}$
and $Q \propto r_{\mathrm{h}}^{-4}$) and anchored to gas production from \citet{Cordiner2025}
and dust production from \citet{Jewitt2025}. Solid curves represent maximum dust-loss
scenarios (120~kg\,s$^{-1}$ at 3.83~AU), while dashed curves represent minimum dust-loss
scenarios (12~kg\,s$^{-1}$ at 3.83~AU). Two nucleus radii are considered: 500~m and
2200~m.
}
\label{fig:erosion}
\end{figure}

\subsection{Implications for Outgassing Material} \label{sec:implications}

Prior to perihelion, the cumulative erosion depth of 3I/ATLAS remains well below the
$\sim$15--20~m depth to which GCR are expected to chemically and structurally modify
cometary material over $\sim$1~Gyr of exposure (see Figures~2 and~3). This indicates that
the volatiles released during the inbound leg of the orbit originate predominantly from
GCR-processed surface layers. Post-perihelion, this conclusion remains valid for larger
nucleus sizes, independent of the assumed heliocentric dependence of the activity. However,
for small-nucleus scenarios (radii of a few hundred meters) combined with steep heliocentric
activity laws (e.g., $Q \propto r_{\mathrm{h}}^{-4}$), the erosion depth can exceed the
cosmic-ray penetration threshold. In such cases, post-perihelion outgassing may begin to
expose and release unprocessed, pristine material from the interior.

A close passage near a star, either the parent star prior to ejection or another star
encountered during interstellar transit, represents a potential scenario that could result
in 3I/ATLAS outgassing pristine material. In this case, thermal processing induced by stellar
radiation could have sublimated and removed pre-existing irradiated surface layers formed by
long-term GCR exposure, thereby exposing deeper, unaltered ices. A collision could also
redistribute the irradiated material \citep[see the discussion in][]{Maggiolo2020}. If the
collision is catastrophic, the nucleus material is altered, and the material within a body
that would form from the reaggregation of the fragments could not be considered pristine
anymore. In the case of a sub-catastrophic or marginally catastrophic collision, the nucleus
material (both processed by GCR and not processed by GCR) is mostly unaltered. The fragments
resulting from such a collision could reaggregate to form a new nucleus and irradiated
material may end up deep inside the newly formed nucleus. If collision debris are large
enough and diffusion in the reformed nucleus limited, this could be a source of nucleus
inhomogeneity.

However, both the stellar-encounter and collision scenarios would lead to the release of
pristine material only if they occurred recently enough to prevent the reformation of a thick
irradiated crust. Just below the surface, where GCR energy deposition is highest, the
timescale to deplete CO is on the order of 500~Myr \citep{Maggiolo2020}. The conversion of
low-temperature porous amorphous ice or of crystalline ice into compact amorphous ice is
quite efficient \citep{Dartois2013,Dartois2015}. In the near-surface region where GCR energy
deposition peaks, structural modifications of the ice occur on relatively short timescales.
At a depth of $\sim$2~m, the transition from crystalline to amorphous ice proceeds over
$\sim$60~Myr, while amorphous ice compacts over $\sim$15~Myr \citep{Maggiolo2020}. As shown
in Figures~2 and~3, 1~Gyr of GCR irradiation is sufficient to produce a chemically and
structurally altered layer tens of meters thick.

In addition to chemical processing, physical erosion due to cosmic-ray-induced sputtering has
also been considered. However, laboratory-based sputtering yields indicate that, for a
typical interstellar GCR ionization rate of $\sim 10^{-16}$~s$^{-1}$, sputtering fluxes range
from $\sim$1 to $10^{3}$~molecules~cm$^{-2}$~s$^{-1}$ for common ices such as H$_2$O, CO$_2$,
and CO \citep{Dartois2023}. Integrated over 1~Gyr, this corresponds to the loss of
$\sim 3.2 \times 10^{16}$ to $3.2 \times 10^{19}$~molecules~cm$^{-2}$, equivalent to erosion of
only $\sim$0.3~$\mu$m to 0.3~mm of surface material (assuming $\sim 10^{22}$~molecules~cm$^{-3}$).
This confirms earlier conclusions that GCR-induced sputtering does not significantly
contribute to surface removal over Gyr timescales and therefore cannot account for the
exposure of unirradiated subsurface layers.

Taken together, these considerations support the conclusion that the volatiles observed in
3I/ATLAS prior to perihelion are consistent with release from GCR-processed material, unless
a recent resurfacing event, such as a close stellar encounter or collision, removed or
redistributed the irradiated mantle and exposed deeper, pristine ices.

\section{Implications and paradigm shift} 

Rather than being pristine messengers from distant planetary systems, interstellar objects
may instead carry signatures of processed material shaped by Gyr-scale cosmic-ray exposure
both in the distant reservoirs of their parent stellar systems, prior to ejection, and
during their interstellar transit to the solar system.

\subsection{3I/ATLAS: From Pristine to Processed} \label{sec:processed}

It is often assumed that interstellar objects retain the primordial compositions acquired in
their formation environments \citep{Jewitt2023}. 

However, beyond any storage within their parent stellar systems, interstellar objects can reside in the dalactic environment for \emph{Gyr-scale} times, as inferred for the first two interlopers and particularly for 3I/ATLAS \citep{Jewitt2023,Taylor2025}. Their surface layers have been exposed to cosmic-ray irradiation for sufficiently long periods to
drive extensive alteration of their outer layers.
 The processed zone, estimated to extend
15--20~m deep, surpasses pre-perihelion erosion depths, implying that pre-perihelion
observations generally probe altered rather than pristine material. Because most erosion
occurs near perihelion, some of these objects may only expose unprocessed interior material
after perihelion, depending on their orbital dynamics, outgassing rates, and sizes.

This reinterpretation fundamentally alters how we interpret spectroscopic measurements of
3I/ATLAS. Rather than sampling primordial material that reveals distant formation
conditions, we are observing the products of Gyr-scale galactic cosmic-ray chemistry. The
extreme CO$_2$/H$_2$O ratio reflects processing efficiency, not formation environment. The
substantial CO levels indicate ongoing chemical evolution, not pristine composition. This
study suggests that 3I/ATLAS is better understood as a natural laboratory for cosmic-ray
processing than as a direct messenger from a distant protoplanetary disk. This concept is
summarized in Figure~5, which illustrates the stratigraphy of an irradiated nucleus, with a
GCR-processed crust overlying a pristine interior shielded from cosmic rays.

\begin{figure}[ht!]
\centering
\includegraphics[width=\textwidth]{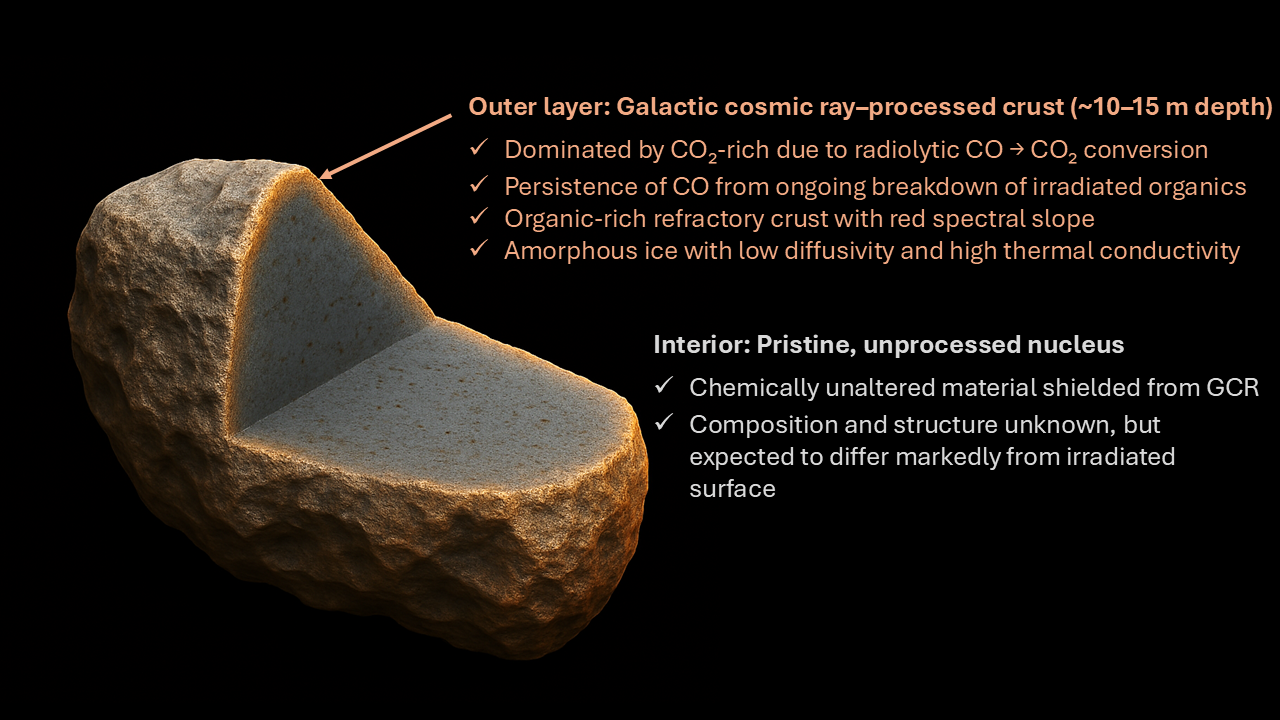}
\caption{
Schematic illustration of interstellar comet 3I/ATLAS showing the expected stratigraphy of
an irradiated nucleus. Galactic cosmic ray irradiation over several Gyr alters the
outer $\sim$15--20~m, producing a CO$_2$-rich crust through CO-to-CO$_2$ conversion, ongoing
CO release from the breakdown of irradiated organics, an organic-rich refractory mantle
with a red spectral slope, and compact amorphous ice characterized by low diffusivity and
high thermal conductivity. Beneath this layer lies a pristine, unprocessed interior shielded
from GCRs, whose composition and structure remain unknown but are expected to differ
markedly from the irradiated crust.
}
\label{fig:atlas_stratigraphy}
\end{figure}

\subsection{Future Observational Strategy for Interstellar Objects} \label{sec:future}

This paradigm shift defines clear observational priorities. For 3I/ATLAS, key tests include
\textit{JWST}/MIRI searches for organic features diagnostic of irradiation processing,
thermal mapping to distinguish altered surface layers from low-conductivity insulating
mantles, and continued monitoring of volatile ratios, in particular for the main products
of H$_2$O and CO$_2$ ice radiolysis such as CO$_2$, O$_2$, H$_2$, and H$_2$O$_2$. Comparing
pre- and post-perihelion activity is especially valuable, since post-perihelion outgassing
may probe deeper layers less affected by GCR irradiation.

Looking ahead, the Vera~C.~Rubin Observatory could detect up to 70 interstellar objects
annually \citep{Marceta2023}, enabling statistical tests of how widespread cosmic-ray
processing is among such bodies. Rather than attributing compositional diversity solely to
differences in formation environments, population studies must account for systematic ageing
effects linked to interstellar residence times.

\section{Conclusion} \label{sec:cite}

Multi-instrument observations of interstellar comet 3I/ATLAS reveal an unprecedented
CO$_2$/H$_2$O ratio of $7.6 \pm 0.3$ \citep{Cordiner2025}, the most extreme volatile
composition ever recorded in a comet. This value lies $4.5\,\sigma$ above solar system
trends and exceeds the typical cometary median of 0.12 reported by
\citet{HarringtonPinto2022} for 25 solar system comets by a factor of $\sim 6763$,
requiring explanation through evolutionary processing rather than primordial inheritance.

The combined evidence, including extreme CO$_2$ enrichment, sustained CO and H$_2$O
outgassing, and red spectral slopes (16--27\%\,per\,k\AA), points toward Gyr-scale galactic
cosmic ray processing as the dominant mechanism shaping 3I/ATLAS’s current
properties. Laboratory irradiation experiments demonstrate that GCRs efficiently convert CO
into CO$_2$ while producing organics and compact amorphous ice
\citep{Hudson1999,Palumbo1997}. Radiation transport models show that such processing alters
the nucleus down to depths of 15--20~m \citep{Gronoff2020,Maggiolo2020}. Estimations of
3I/ATLAS erosion confirm that present-day activity exposes only this GCR-processed crust,
while the pristine interior remains shielded.

This reinterpretation represents a paradigm shift: interstellar objects cannot be assumed
to be pristine messengers of their formation environments. Instead, they serve as natural
laboratories of long-term cosmic-ray chemistry.

With perihelion in October~2025, time-sensitive observations remain crucial.
\textit{JWST}/MIRI searches for organic features, thermal mapping of insulating crusts, and
continued volatile monitoring can critically constrain the extent of GCR-induced
alteration.

\begin{acknowledgments}
JDK and FD are supported by ESA/PRODEX Comet Interceptor under contract PEA 4000139830

\begin{contribution}

R. M. came up with the initial research concept, participated to the development and design of methodology; participated to the creation of models; edited the manuscript and was responsible for writing and submitting the manuscript. F. D. participated to the preparation, creation and presentation of the published work, specifically writing the initial draft. G. G. came up with the initial research concept; participated to the development and design of methodology and participated to the creation of models. J. de K. and G. C. participated to the preparation, creation and  presentation of the published work.


\end{contribution}

\bibliography{references_apj_final}{}
\bibliographystyle{aasjournal}



\end{acknowledgments}
\end{document}